\documentclass[conference]{IEEEtran}
\IEEEoverridecommandlockouts
\usepackage{cite}
\usepackage{amsmath,amssymb,amsfonts}
\usepackage{algorithmic}
\usepackage{graphicx}
\usepackage{textcomp}
\usepackage{xcolor}
\def\BibTeX{{\rm B\kern-.05em{\sc i\kern-.025em b}\kern-.08em
    T\kern-.1667em\lower.7ex\hbox{E}\kern-.125emX}}
\begin{document}

\title{Poster: ARLIF-IDS --- Attention augmented Real-Time Isolation Forest Intrusion Detection System}

\author{\IEEEauthorblockN{ Aman Priyanshu}
\IEEEauthorblockA{\textit{Dept. of Information \& } \\
\textit{Communication Technology}\\
\textit{Manipal Institute of Technology}\\
Manipal, India \\
aman.priyanshu@learner.manipal.edu}
\and
\IEEEauthorblockN{ Sarthak Shastri}
\IEEEauthorblockA{\textit{Dept. of Information \& } \\
\textit{Communication Technology}\\
\textit{Manipal Institute of Technology}\\
Manipal, India \\
sarthak.shastri@learner.manipal.edu}
\and
\IEEEauthorblockN{Sai Sravan Medicherla}
\IEEEauthorblockA{\textit{Dept. of Information \& } \\
\textit{Communication Technology}\\
\textit{Manipal Institute of Technology}\\
Manipal, India \\
sai.medicherla@learner.manipal.edu}
}

\maketitle

\begin{abstract}
Distributed Denial of Service (DDoS) attack is a malicious attempt to disrupt the normal traffic of a targeted server, service or network by overwhelming the target or its surrounding infrastructure with a flood of Internet traffic. Emerging technologies such as the Internet of Things and Software Defined Networking leverage lightweight strategies for the early detection of DDoS attacks. Previous literature demonstrates the utility of lower number of significant features for intrusion detection \cite{josyDDOS, Aburomman, fitni}. Thus, it is essential to have a fast and effective security identification model based on low number of features. 

In this work, a novel Attention-based Isolation Forest Intrusion Detection System is proposed. The model considerably reduces training time and memory consumption of the generated model. For performance assessment, the model is assessed over two benchmark datasets, the NSL-KDD dataset \& the KDDCUP'99 dataset. Experimental results demonstrate that the proposed attention augmented model achieves a significant reduction in execution time, by 91.78\%, and an average detection F1-Score of 0.93 on the NSL-KDD and KDDCUP'99 dataset. The results of performance evaluation show that the proposed methodology has low complexity and requires less processing time and computational resources, outperforming other current IDS based on machine learning algorithms.
\end{abstract}


\section{Introduction}
The rapid advances in the communication technology have resulted in transmission services handling huge network size and the corresponding data. As a result, malicious entities have ample opportunity to attack and deny service to target systems (DDoS). Despite existing traditional solutions, Distributed Denial of Service (DDoS) attacks continue to be a prominent cybercrime. Many attacks occur each day and pose security challenges for client-networks. Therefore, the existence of accurate intrusion detection systems becomes indispensable. An intrusion detection system (IDS) is a tool that prevents the network from possible intrusions by inspecting the network traffic, to ensure its confidentiality, integrity, and availability. Despite enormous efforts by the researchers, IDS still faces challenges in improving detection performance while reducing detection-time and increasing memory optimization.

\section{Related Work}

Existing surveys on anomaly detection techniques, only marginally consider the utility of neural-network-based approaches, due to their intensive memory consumption and high detection-time. Current research developments regarding deep neural networks and LSTM architectures for anomaly detection are oftentimes not scalable due to these issues. However, IDS which employ machine learning methodology, such as classification-based, clustering-based and information-theoretic approaches. Thus, techniques based on the  isolation forest (IF), random forest (RF), distance-based inference, entropy-based inference constitute a major part of the investigations \cite{josyDDOS}. Theoretical studies have also been conducted on LSTM and CNNs. An analysis is conducted over these deep neural network methods.

\section{Methodology}

Our proposed methodology, which we call ARLIF - Attention augmented Real-Time Isolation Forest Intrusion Detection System, draws inspiration from the transformer models \cite{attention}. However, unlike traditional attention-based model, we employ learning on only the attention layer, whereas time-series datum is computed using proba values predicted by Isolation Trees. The methodology allows online learning, and real-time detection systems may employ it for high-performance detection.

The architecture of this model relies solely on attention mechanisms to carry out the seasonality learning, allowing the modulation and focus of predictions to be placed on specific isolation trees, based on previous responses (\textit{context}).

\subsection{Attention Mechanism}

\begin{figure*}[t]
\centerline{\includegraphics[height=75mm]{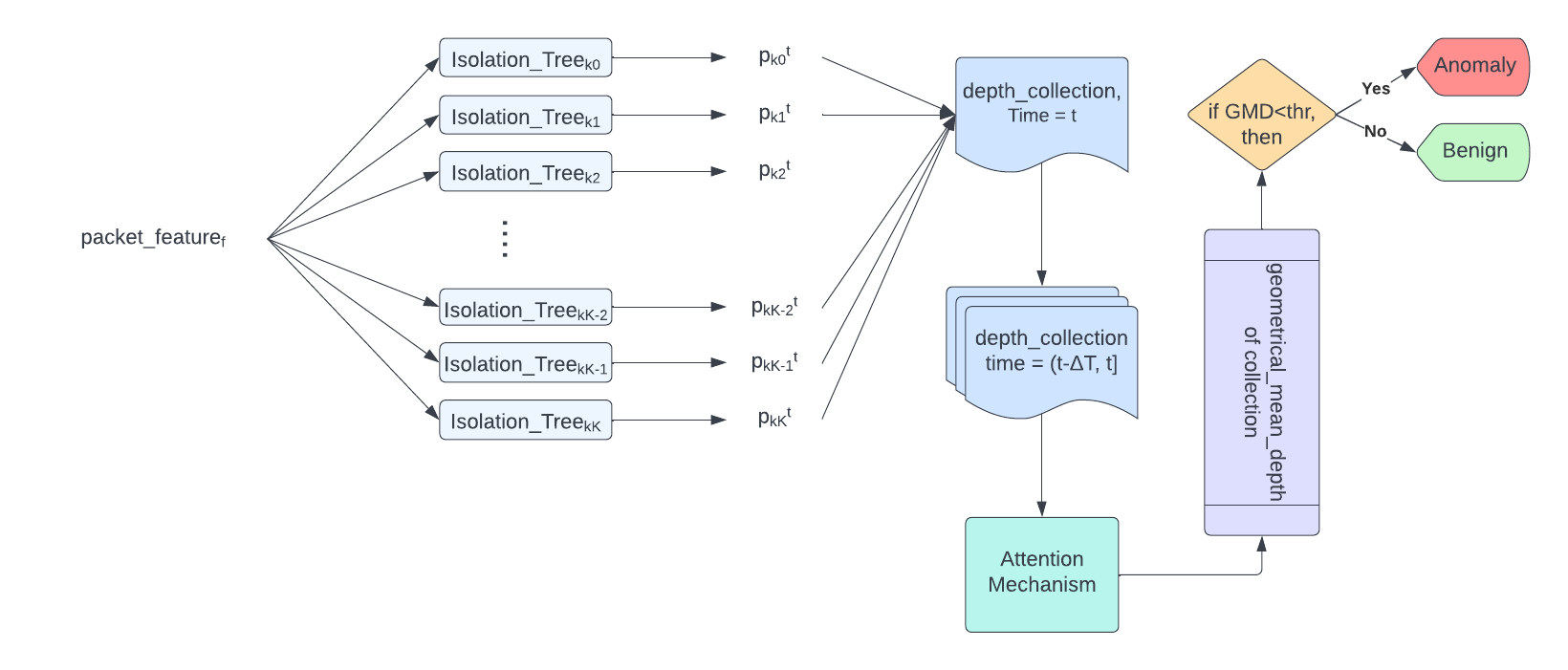}}
\caption{Proposed methodology - Attention augmented Real-Time Isolation Forest Intrusion Detection System}
\label{fig:1}
\end{figure*}

In order to utilize the attention mechanism in the network intrusion detection domain and take advantage of its properties, we employ its contextualizing attentive properties during forest response averaging. Taking each trees predicted probability as its vector at $V_{k}^{i}$ for $tree^{i}$ at $time=k$. Thereby, allowing the attention mechanism to create updated vector embeddings for every tree, based on responses from others as well as, their time-series.

To formulate the algorithm presented in Fig~\ref{fig:1}, we limit previous time-series predictions of each tree, to be previous $k$-responses. Therefore once a forest of isolation trees are trained, the online learning will only be executed through the attention mechanism, which would have: $number\_of\_trainable\_parameters = 3 * k * (k+1)$, which would be based on $query, key, \& value$ layers.
The model therefore, offers a memory optimal, time efficient approach towards intrusion detection.

\section{Results}

\begin{table}[h]
\caption{Model Performance Inference on Datasets}
\begin{center}
\begin{tabular}{|c|c|c|c|}
\hline
\textbf{Model Name} & \textbf{F1-Score} & \textbf{Memory Acquired} & \textbf{Detection-Time} \\
\hline
CNN & 0.95 & 105kB & 4.6s\\
\hline
LSTM & 0.96 & 238kB & 7.3s\\
\hline
ARLIF-IDS & 0.93 & 11kB & 0.6s\\
\hline
Isolation Forest & 0.87 & 6kB & 0.4s\\
\hline
Random Forest & 0.88 & 44kB & 0.9s\\
\hline
Entropy-based IDS & 0.83 & 3kB & 0.2s\\
\hline
Distance-based IDS & 0.84 & 4kB & 0.2s\\
\hline
\end{tabular}
\label{table:1}
\end{center}
\end{table}

We carry out a set of experimental evaluations in order to provide a deep analysis of the performance of the proposed model. Then, we compare the overall results against baseline models. The models are tested on KDDCUP'99 dataset and the NSL-KDD dataset as benchmark.

We present these results in Table~\ref{table:1}. As demonstrated the proposed methodology is able to give high preliminary results for the memory-time-performance tradeoff. The table presents neural network and tree based methods as the highest performing algorithms, where as, entropy-based and distribution-based methods provide low memory and execution-time utility. ARLIF-IDS gives a satisfactory trade-off giving performance even higher than LSTM networks.

\section{Conclusion}

This poster presents an intrusion detection model with attention mechanism for isolation forest augmentation. We take into account the sparse correlation majority traffic-features have with anomaly detection, and develop a strategy to utilize models with low-feature space input further cutting down on computational costs \cite{josyDDOS}. The proposed model was demonstrated to achieve satisfactory performance of 0.93 F1-Score for the KDDCUP'99 dataset and NSL-KDD dataset. We also show that it lowers average detection-time of 91.78\% with respect to LSTM anomaly detection model and 95.37\% lower memory utilization with respect to the same. For future work, we hope to utilize other probabilistic methods instead of isolation trees.

\vspace{12pt}

\end{document}